\documentclass[12pt]{iopart}
\usepackage{epsfig}  

\newcommand{\be}{\begin{equation}}
\newcommand{\eq}{\end{equation}}

\def\beqa{\begin{eqnarray}}
\def\eeqa{\end{eqnarray}}

\newcommand{\CA}{{\cal A}}

\newcommand{\CN}{{\cal N}}
\newcommand{\CC}{{\cal C}}

\newcommand{\CP}{{\cal P}}

\newcommand{\CZ}{{\cal Z}}
\newcommand{\CO}{{\cal O}}

\newcommand{\CW}{{\cal W}}

\begin{document}
\jl{6}

\begin{center}
\hfill AEI-1999-38\\[1mm]
\hfill {\tt hep-th/9911170}\\
\end{center}

\title{Yang-Mills Integrals }

\author{Werner Krauth\ddag, Jan Plefka\S\ and Matthias 
Staudacher\S\footnote[5]{Talk
presented by M.~Staudacher at Strings '99, Potsdam, July 19-24 1999}
}

\address{\ddag 
CNRS-Laboratoire de Physique Statistique,
Ecole Normale Sup\'{e}rieure,
24, rue Lhomond, F-75231 Paris Cedex 05, France}
\address{\S
Max-Planck-Institut f\"ur Gravitationsphysik,
Albert-Einstein-Institut,
Am M\"uhlenberg 1, D-14476 Golm, Germany}

\begin{abstract}
$SU(N)$ Yang-Mills integrals form a new class of matrix models which, 
in their maximally supersymmetric version, are relevant to recent
non-perturbative definitions of  10-dimensional IIB superstring
theory and 11-dimensional M-theory. We demonstrate how Monte Carlo
methods may be used to establish
important properties of these models. In particular we consider 
the partition functions as well as 
the matrix eigenvalue distributions. For the latter we derive a number
of new exact results for $SU(2)$. 
We also report preliminary computations of Wilson loops.

\end{abstract}


\submitted

\maketitle

\section{Motivation}

Recently there has been considerable interest in dimensionally
reduced Yang-Mills theories as a means to obtain non-perturbative
information on superstring theory and M-theory. 
The possible relevance of these systems to quantum gravity
appeared in \cite{dhn} through the light cone quantization of
the 11-dimensional supermembrane. It was argued in \cite{dhn}
that 10-dimensional $SU(N)$ super Yang-Mills theory reduced to one
dimension -- i.e.~matrix quantum mechanics -- correctly
quantizes the supermembrane in the large $N$ limit. The very same 
system, first studied (without reference to applications to
quantum gravity) in \cite{ch}, has recently been 
interpreted as a non-perturbative attempt at
M-theory \cite{bfss}. Unfortunately, on the technical side, very
little is known about this model: It is suspected \cite{dhn},
\cite{bfss} that a novel, intricate large $N$ limit is required,
but only few concrete results are available. This motivates the 
study of a simpler system: The complete reduction of Yang-Mills
theory to $0+0$ dimensions. In addition, the reduction of the
ten-dimensional susy gauge theory path integral to a matrix integral
has been at the heart of an alternative proposal to directly define
non-perturbative IIB string theory \cite{ikkt1}. More generally we
may study the complete reduction of D-dimensional $SU(N)$ Yang-Mills
theory.  Then the path integral of the field theory simplifies to
an integral over the group's Lie algebra, with a flat measure:
a {\it Yang-Mills integral}. Denoting the gauge potential by
$X_\mu^A$ and their superpartners by $\Psi_{\alpha}^A$ we obtain

\begin{equation}
\CZ_{D,N}^{\CN}:=\int \prod_{A=1}^{N^2-1} 
\Bigg( \prod_{\mu=1}^{D} \frac{d X_{\mu}^{A}}{\sqrt{2 \pi}} \Bigg) 
\Bigg( \prod_{\alpha=1}^{\CN} d\Psi_{\alpha}^{A} \Bigg)
\exp \bigg[  -S(X,\Psi )
\bigg].
\label{susyint}
\end{equation}
with the Euclidean ``action''
\begin{equation}
S( X,\Psi ) =  -\frac{1}{2} \Tr 
[X_\mu,X_\nu] [X_\mu,X_\nu] - 
\Tr \Psi_{\alpha} [ \Gamma_{\alpha \beta}^{\mu} X_{\mu},\Psi_{\beta}].
\label{action}
\end{equation}
The {\it a priori}
allowed dimensions for the reduced supersymmetric gauge theory are
$D=3,4,6,10$ corresponding to $\CN=2,4,8,16$ real supersymmetries.
For the bosonic, i.e.~the non-supersymmetric 
case $\CN=0$, we omit the Grassmann variables $\Psi_{\alpha}^A$,
and may study all dimensions $D \geq 2$.

There are numerous further reasons for being interested in the
integrals eq.(\ref{susyint}):
 
$\bullet$ The susy integrals are crucial for the computation
of the Witten index of the above mentioned quantum mechanical 
gauge theories, as they contribute to the so-called bulk part of the index
\cite{sestern} (cf also \cite{smilga} for an earlier
calculation). 

$\bullet$ In the
maximally supersymmetric case, the system
describes to leading order the statistical
distribution of a system of $N$ D-instantons (or ``$-1$-branes'').

$\bullet$ The $\CN=16$ integral appears in very recent work developing
a multi-instanton 
calculus for the $\CN=4$, $D=4$ $SU(\infty)$ conformal gauge theory
\cite{hollo1}, and again in the large $N$ limit of Sp$(N)$ and
SO($N$) $\CN=4$ susy gauge theory \cite{hollo2}.

$\bullet$ Finally we can regard the integrals eq.(\ref{susyint}) as
a version of the Eguchi-Kawai reduced gauge theory. The original
work \cite{ek} focused on a lattice formulation and employed
unitary matrices, while the above integrals use 
the hermitean 
gauge connections $X_\mu^A$. This is similar to \cite{gk}; however,
we apply neither gauge fixing nor quenching prescriptions to
the above integrals. The interesting question is whether the models
eq.(\ref{susyint}) encode universal information on the full gauge
field theory as $N \rightarrow \infty$.

The integrals of eq.(\ref{susyint}) appear to be singular
due to the ``valleys'' of the action, i.e.~the directions in the
configuration space of the $X_\mu^A$ where all $D$ matrices commute.
Recent work has however proven this intuition wrong: 
Yang-Mills integrals
{\it do} exist in many interesting cases. 

\section{Partition functions: Convergence properties and 
some exact results}

Indeed, the rigorous results of \cite{sestern} 
(see also \cite{smilga}) 
show that for the gauge group $SU(2)$ the susy integrals converge 
in dimensions $D=4,6,10$. The calculations are easily repeated
for the bosonic case \cite{kns}, and the convergence condition
$D \geq 5$ is found. Unfortunately, no rigorous methods
exist to date for higher rank gauge groups $N \geq 3$. 

In \cite{kns},
\cite{ks1},\cite{ks2} we developed methods to numerically test convergence
of singular multidimensional integrals. The idea is to perform a 
Metropolis random walk weighted by the integrand, and to merely measure
the autocorrelation function of subsequent configurations.  In this
approach, a unit autocorrelation function signals the presence of
a nonintegrable singularity.

As an illustration we plot in Fig.1 the autocorrelation function of the 
$SU(2)$ bosonic integrals. 
\begin{center}
\begin{picture}(200,180)
\put(-300,-120){ \includegraphics{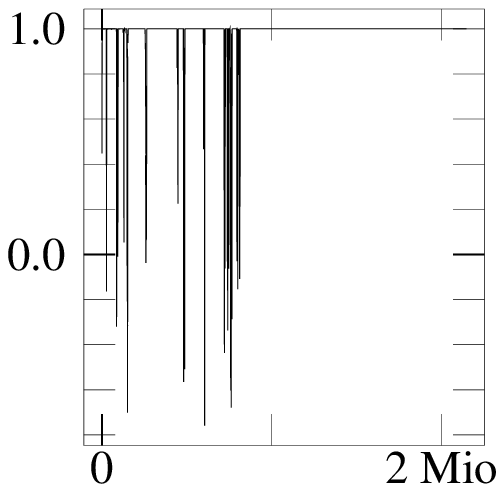} }
\put(40,-120){ \includegraphics{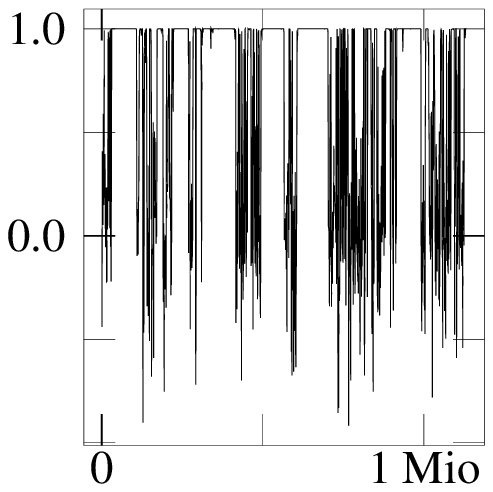} }
\put(360,-120){ \includegraphics{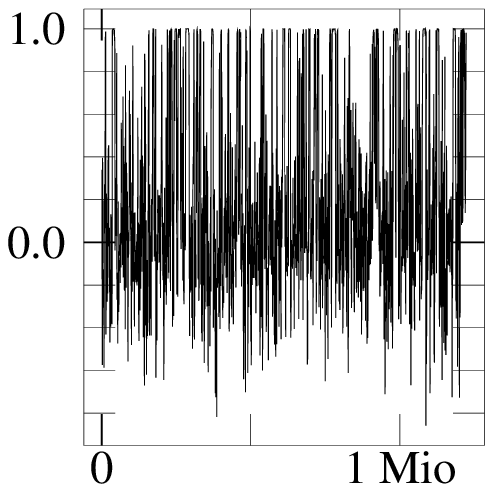} }
\end{picture}
\end{center}
Fig. 1 {\it Autocorrelation functions 
versus Monte Carlo time 
for the $SU(2)$
bosonic integral with, from the left, $D=3,4,5$.\\}

We are clearly able to reproduce the convergence
condition $D \geq 5$: For $D=5$ the configurations decorrelate well and
the whole integration space is properly sampled. (One observes
increasingly improved decorrelation for $D=6,7, \ldots$, not shown
in Fig.1.) In contrast for $D=3$ the system gets quickly trapped in 
a singular configuration:
The Markov chain gets lost in a valley, and the integral is divergent.
$D=4$ shows marginal
divergence, which agrees with the exact analytical results.

Applying the same method to higher rank bosonic models, and to the
supersymmetric models, we are able to map out the convergence conditions
for the Yang-Mills integrals. They read
\begin{eqnarray} 
\left.
\begin{array}{ccc}
D=4,6,10 & \quad {\rm and} \quad &  N \geq 2 
\end{array}
\right\} 
&\quad {\rm for} \;\;\; & \CN >0\cr 
 & & \cr
\left.
\begin{array}{ccc}
\quad D=3 & \qquad {\rm and} \quad & \;N \geq 4 \\
& & \\
\quad D=4 & \qquad {\rm and} \quad & \;N \geq 3 \\
& & \\
\quad D\geq5 & \qquad {\rm and} \quad &\;N \geq 2 
\end{array}  
\right\} 
&\quad {\rm for}\;\;\;& \CN=0
\label{intcond}
\end{eqnarray}
In particular the $D=3$ susy integral is divergent (see
\cite{kns},\cite{ks1},\cite{ks2} for a more detailed discussion
of this point).

It would be nice to have a rigorous mathematical proof of the
conditions (\ref{intcond}). Some understanding may be gained
by considering one-loop perturbative estimates of the integrals
eq.(\ref{susyint}). One has for the supersymmetric case \cite{ikkt2}
\begin{eqnarray}
\CZ_{D,N}^{\CN} \sim 
\int \prod_{i,\mu}^{N,D} dx_\mu^i&~\Bigg[ \prod_\mu \delta\Big( 
\sum_i x_\mu^i \Big) \Bigg]~\times \nonumber\\
&\times~\sum_{G:\;{\rm maximal \; tree}}~
\prod_{(ij):\;{\rm link \; of }\; G}~{1 \over (x^i-x^j)^{3 (D-2)}} + \ldots
\label{susyfluct}
\end{eqnarray}
where the $x_\mu^i$ are the diagonal components of the matrices
$X_\mu^A$, 
while it is found in \cite{nishi} that the bosonic integrals are
approximated by
\begin{equation}
\CZ_{D,N}^{\CN=0} \sim \int \prod_{i,\mu}^{N,D} dx_\mu^i~\Bigg[ \prod_\mu 
\delta\Big( \sum_i x_\mu^i \Big) \Bigg]~ 
\prod_{i<j}~{1 \over (x^i-x^j)^{2 (D-2)}}. 
\label{fluct}
\end{equation}
Powercounting for large separations $(x^i-x^j)^2$ yields
precisely eq.(\ref{intcond}). However, it should be stressed 
that this does not prove the convergence conditions, since one
has to worry about configurations where some 
separations are small, so that the one-loop approximation becomes 
invalid.

In the supersymmetric case the value for the integrals is
believed to be known:
\begin{equation}
\CZ_{D,N}^{\CN}=\frac
{2^{\frac{N(N+1)}{2}} \pi^{\frac{N-1}{2}}}
{2 \sqrt{N} \prod_{i=1}^{N-1} i!} 
\left\{ 
\begin{array}{ccc}
\frac{1}{N^2} & D=4, & \CN=4 \\ 
&\\
\frac{1}{N^2} & D=6, & \CN=8 \\
&\\ 
\sum_{m | N} \frac{1}{m^2} & D=10, & \CN=16
\end{array} \right.
\label{index}
\end{equation}

The $\CN=16$ expression was conjectured by Green and Gutperle \cite{greengut}
based on a calculation of the D-instanton effective action of the
superstring. A derivation of the terms to the right
of the curly bracket in eq.(\ref{index}),
based on cohomological deformation techniques,
was given in \cite{moore}. This calculation has still
an important loophole (see comments in \cite{ks2}).
However, the formula eq.(\ref{index}) was numerically checked by
Monte Carlo techniques in \cite{kns},\cite{ks1} up to $N \sim 5$.

For the bosonic case no exact value of the partition function is
known except for $SU(2)$ \cite{kns}, where the result reads
\begin{equation}
\CZ_{D,2}=
2^{-\frac{3}{4}D -1} \frac{\Gamma(\frac{D}{4}) \Gamma(\frac{D-2}{4})
\Gamma(\frac{D-4}{4})}{\Gamma(\frac{D}{2}) \Gamma(\frac{D-1}{2})
\Gamma(\frac{D-2}{2})}  \qquad {\rm for} \qquad D \geq 5
\label{nosusy}
\end{equation}

It would be exciting to find the generalization of this result
to higher rank gauge groups -- this is after all the ``zero-mode''
contribution to the Yang-Mills partition function on a D-dimensional
torus.  

\section{Eigenvalue densities: Asymptotics and the exact
$SU(2)$ densities}

Let us shift attention from the partition functions to the correlation
functions of the models. The simplest correlators are $SU(N)$ invariant
one-matrix correlators: the moments $\langle \Tr X_D^k \rangle$ of one
matrix, say the $D$-th: $X_D$. 
These are directly related to the distribution of eigenvalues
of the matrix: If the eigenvalues of $X_D$ are $\lambda_1, \ldots, \lambda_N$,
the eigenvalue density is defined for all $N$ as
\begin{equation}
\rho(\lambda)=\langle \frac{1}{N} \sum_{i=1}^N \delta(\lambda-\lambda_i)
\rangle.
\label{density}
\end{equation}
The non-zero moments of $\rho(\lambda)$ are then given by
\begin{equation}
\Big\langle {1 \over N}\Tr X_D^{2 k} \Big\rangle = 
\int_{-\infty}^{\infty} d\lambda \rho(\lambda)~\lambda^{2 k}
\label{corr}
\end{equation}
In an ordinary Wigner type matrix model all moments exist. Yang-Mills
integrals are more intricate. Analytical calculations have only 
been performed for $SU(2)$, and we found the following surprising
results. In the $D=4$ susy integral all moments are infinite, even
though the integral itself exists, as argued above. In the $D=6$ susy
integral the first two moments are finite and one finds
\begin{eqnarray}
\langle \Tr X_D^2\rangle_{D=6} = \frac{1}{2}\,\sqrt{\frac{2}{\pi}}&\qquad&
\langle \Tr X_D^4\rangle_{D=6} = \frac{25}{64}
\end{eqnarray}
while all higher moments diverge. For the $D=10$ susy integral we have 
exactly twelve
finite moments which are
\begin{eqnarray}
\langle \Tr X_D^2\rangle_{D=10} = \frac{8}{25}\,\sqrt{\frac{2}{\pi}}
&\qquad&
\langle \Tr X_D^4\rangle_{D=10} = \frac{9}{80}\nonumber\\
\langle \Tr X_D^6\rangle_{D=10} = \frac{3}{32}\,\sqrt{\frac{2}{\pi}}
&\qquad&
\langle \Tr X_D^8\rangle_{D=10} = \frac{297}{4096}\nonumber\\
\langle \Tr X_D^{10}\rangle_{D=10} = \frac{1089}{8192}
\,\sqrt{\frac{2}{\pi}}&\qquad&
\langle \Tr X_D^{12}\rangle_{D=10} = \frac{184041}{655360}
\end{eqnarray}
It would be interesting to find a geometrical or combinatorial interpretation
for these numbers. Which densities give rise to this convergence behavior?
For $SU(2)$ we can go farther and find the exact densities:
\begin{eqnarray}
\rho_{D=4}^{\rm{SUSY}}(\lambda)&=& \frac{3\cdot 2^{5/4}}{\sqrt{\pi}}
\, \lambda^2\, U(\frac{5}{4}, \frac{1}{2} , 8\lambda^4)
\label{exactrho}\\ 
\rho_{D=6}^{\rm{SUSY}}(\lambda)&=& \frac{105}{2^{3/4}\, \sqrt{\pi}}
\, \lambda^2\, \Bigl [ U(\frac{9}{4}, \frac{1}{2} , 8\lambda^4) -
\frac{33}{16}\, U(\frac{13}{4}, \frac{1}{2} , 8\lambda^4)\, \Bigr ] 
\nonumber\\
\rho_{D=10}^{\rm{SUSY}}(\lambda)&=& \frac{1287}{64\cdot 2^{3/4}\, \sqrt{\pi}}
\, \lambda^2\, \Bigl [ 546\,  U(\frac{17}{4}, \frac{1}{2} , 8\lambda^4)
\nonumber \\
&& \qquad \qquad \qquad
-147\, \frac{17\cdot 19}{8}\,  U(\frac{21}{4}, \frac{1}{2} , 8\lambda^4)
\nonumber\\
&& \qquad \qquad \qquad + 45\, \frac{17\cdot 19\cdot 21\cdot 23}{256}\, 
 U(\frac{25}{4}, \frac{1}{2} , 8\lambda^4) \nonumber \\
&& \qquad \qquad \qquad 
-\frac{17\cdot 19\cdot 21\cdot 23\cdot 25\cdot 27}{2048}\,
 U(\frac{29}{4}, \frac{1}{2} , 8\lambda^4)\, \Bigl ]\nonumber
\end{eqnarray}
where $U$ is the Kummer-$U$ function defined as
\begin{equation}
U(a,b,z)= \frac{1}{\Gamma(a)}\,
\int_0^\infty dt \, t^{a-1}~(1+t)^{b-a-1}~e^{-z t}
\label{kummer}
\end{equation}
Now we see that the above finiteness properties of the moments result
from a rather curious powerlike behavior of the densities at large
values of $\lambda$. We have for $\lambda \rightarrow \infty$
\begin{equation}
\rho^{\rm SUSY}_D(\lambda) \sim
\left\{ 
\begin{array}{cc}
\lambda^{-3} & \qquad D=4  \\ 
&\\
\lambda^{-7} & \qquad D=6  \\
&\\ 
\lambda^{-15} & \qquad D=10 
\end{array} \right.
\label{asymp}
\end{equation}
This power-like behavior is very different from Wigner
type systems where the fall-off at infinity is at least exponential.
For the $D$-dimensional bosonic models the density can be worked
out as well albeit less explicitly than in eqs.(\ref{exactrho}), and one
finds the asymptotic behavior  $\rho_D(\lambda) \sim \lambda^{3-D}$.  

Moving on to higher values of $N$, we are unable to analytically
calculate the eigenvalue densities with presently known techniques.
We can, however, find numerically exact densities using Monte Carlo
methods. In Fig.2 we illustrate this by plotting the $N=2,3,4$ $D=4$
susy half-densities (we only plot the $\lambda \geq 0$ part since
the densities are symmetric functions). In the $SU(2)$ case  
the exact expression of eq.(\ref{exactrho}) and the Monte Carlo data
cannot be separated on the scale of the figure.

Now we would like to know how the $SU(2)$ result eq.(\ref{asymp})
generalizes to other values of $N$. It is impossible to extract the
asymptotics 
from histograms such as Fig.2, since the tails comprise only a small
number of samples. Instead, we can go back to the Markov-chain technique
of the last section and measure the finiteness of the moments 
eq.(\ref{corr}) for various values of $N$. We find that for the susy
integrals the qualitative behavior of the $SU(2)$ case persists:
In the $D=4,6,10$ integrals only the first, respectively, $0,2,12$ moments are
finite. We thus conclude, in view of eq.(\ref{corr}), 
that the asymptotic behavior {\it eq.(\ref{asymp}) is
valid for all $N$}. This is very different from Wigner
type random matrix models, where as $N$ increases, the density condenses
onto a compact interval. At the sharply defined (at $N=\infty$)
edge of the interval Wigner distributions show universal behavior.
We have argued that in susy Yang-Mills matrix models no such edge exists,
indicating that {\it the large $N$ physics of these models is indeed very 
different}.

\begin{center}
\begin{picture}(200,180)
\put(-320,220){ \includegraphics{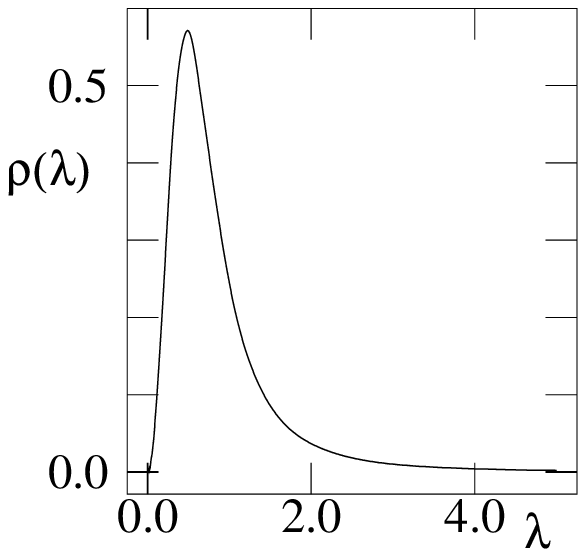} }
\put(20,220){ \includegraphics{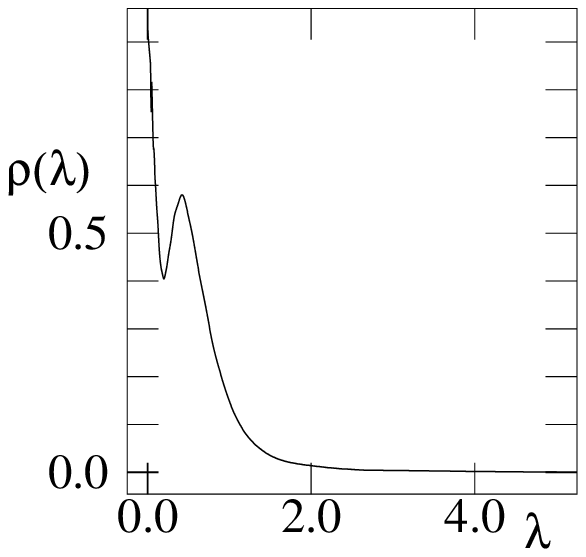} }
\put(360,220){ \includegraphics{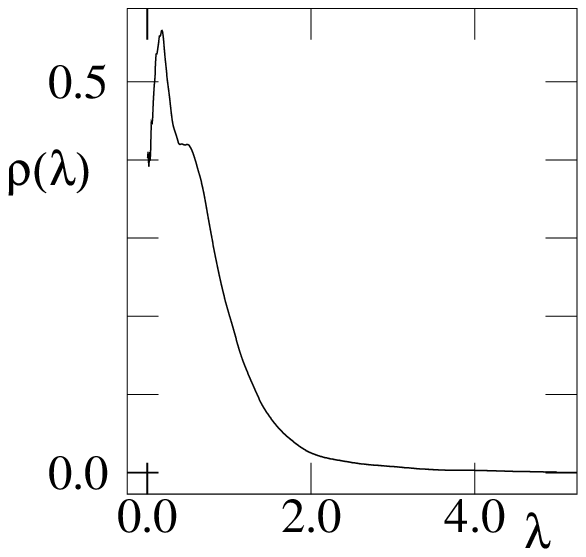} }
\end{picture}
\end{center}
Fig. 2 {\it Eigenvalue (half-)densities for susy $D=4$, 
from the left, $N=2,3,4$\\}

Let us furthermore compare supersymmetric and non-supersymmetric
Yang-Mills integrals. How does the asymptotic behavior of the density
eq.(\ref{asymp}) change in the absence of susy? We have already mentioned
above that for $SU(2)$ this behavior is powerlike as well. Actually
we can guess a general formula by looking once more at the effective
one-loop estimate of eq.(\ref{fluct}). For one-matrix correlators the
most dangerous configuration stems from pulling away one coordinate
$x_D^i$ from a bulk configuration of all other $D-1$ coordinates.
Powercounting leads to the guess
\begin{equation}
\rho_D(\lambda) \sim
\lambda^{-2 N (D-2)+3 D-5} \qquad {\rm where} \qquad
N>\frac{D}{D-2}
\label{bosasymp}
\end{equation}
The same procedure applied to the susy estimate eq.(\ref{susyfluct}) 
reproduces eq.(\ref{asymp}). We then verified the validity of 
eq.(\ref{bosasymp}) by the same Monte Carlo random walk procedure
as above, measuring the finiteness of moments.
We thus notice a marked difference to the susy situation: In the
bosonic case {\it all moments exist} as $N \rightarrow \infty$
for all $D \geq 3$. In particular we expect the eigenvalue distribution
to condense onto a compact support, much like for Wigner type models. 

\section{Wilson loops: Preliminary Results}

A further natural set of correlation functions of
Yang-Mills matrix integrals are Wilson loops. 
Due to the Eguchi-Kawai mechanism \cite{ek},\cite{gk}, 
one {\it naively} expects them
to correspond at $N=\infty$ to Wilson loops in the unreduced
gauge field theory. In the proposal of \cite{ikkt1} for a non-perturbative
definition of the IIB superstring, they have been interpreted as
string creation operators \cite{ikkt3}.

Despite the dimensional
reduction of the field theory to zero dimensions we are still able
to define an infinite 
set of independent Wilson loops dependent on an arbitrary
contour $\CC$ in $D$-dimensional Euclidean space:
\begin{equation}
\CW(\CC)=\langle \frac{1}{N} \CP \Tr e^{i \oint_{\CC} d y_\mu X_\mu} \rangle
\label{wilson}
\end{equation}
Due to the non-commutative nature of the connections $X_\mu$ and
the path-ordering $\CP$, this is a non-trivial functional of
the contour $\CC$ despite the fact that the $X_\mu$ are 
spacially constant.
In the special case of a rectangular contour with lengths $L$ and $T$
in the $(y_1,y_D)$ plane this simplifies to 
\begin{equation}
\CW(L,T)=
\langle \frac{1}{N} \Tr e^{i L X_1} e^{i T X_D} e^{-i L X_1} e^{-i T X_D} 
\rangle
\label{rect}
\end{equation}
We would like to understand how the loops $\CW$ behave as a function
of $N$ and as a functional of the shape of the contours, in particular
whether planar loops satisfy an area-law. We would also like to see
whether there are any telltale differences between the supersymmmetric
and the bosonic loops.

In the previous sections we have shown how a number of exact results
may be derived for the simplest gauge group $SU(2)$. We have not been
able to analytically calculate a Wilson loop for an Yang-Mills
integral even for $SU(2)$. On the other hand it is possible to 
obtain high precision numerical results for low values of $N$.
In Fig.3a we plot the Wilson loop for a square ($L=T$) in the
case of susy $D=4$ as a function of $L$ for various
values of $N$.

\begin{center}
\begin{picture}(200,180)
\put(-200,-40){ \includegraphics{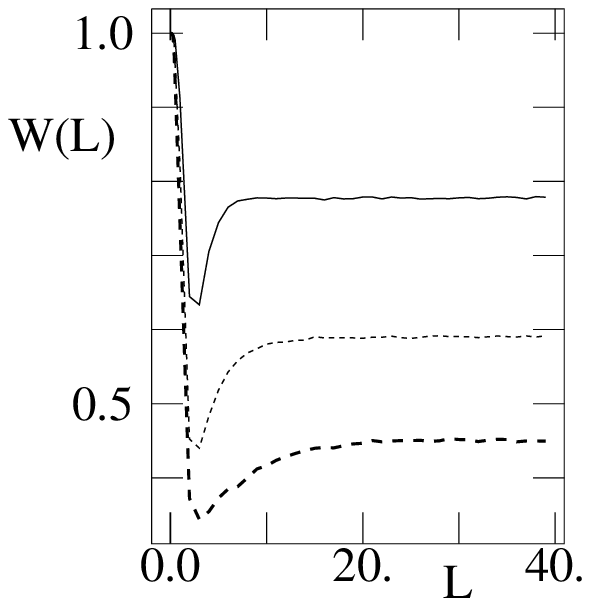} }
\put(200,-40){ \includegraphics{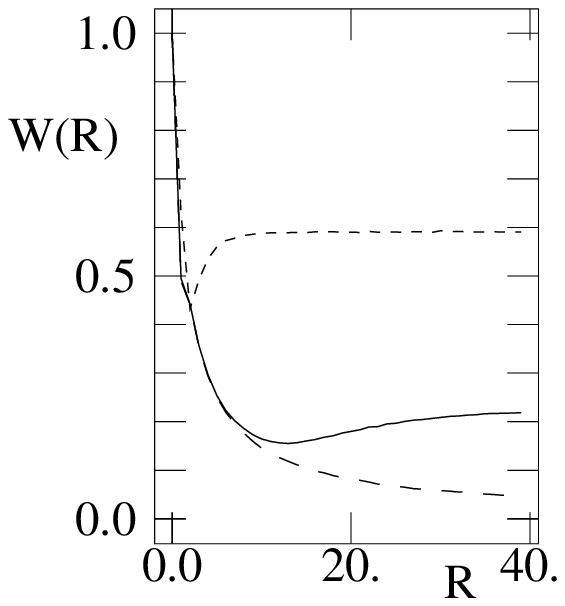} }
\end{picture}
\end{center}
Fig. 3 $D=4$ Susy Wilson loops {\it (a) 
Square of side length $L$ with $N=2,4,8$ (top to bottom)
(b) square, regular 16-gon and 64-gon enclosed in circle 
of radius $R$ with $N=4$.\\}

\
The behavior for small area is easy to understand.
Indeed, for an arbitrary planar loop enclosing a small area of size $\CA$,
it is straightforward to show, using
Stokes' theorem, that
$\CW(\CC)=1+\frac{1}{2 N} \CA^2 \langle \Tr [X_1,X_2]^2 \rangle +
\CO(X^6)$. This immediately gives
\begin{equation}
\CW(\CC)=
\left\{ 
\begin{array}{cc}
1-\frac{1}{4 N} \frac{N^2-1}{D-1} \CA^2 + \ldots & \qquad \CN=0  \\ 
&\\
1-\frac{1}{2 N} \frac{N^2-1}{D} \CA^2 + \ldots & \qquad \CN > 0   
\end{array} \right.
\label{small}
\end{equation} 
and agrees to high precision with the numerical data. A very curious
feature of Fig.3a is that the loops tend to a constant for large area.
The existence of this constant can be demonstrated analytically. It should 
be considered a finite $N$ artifact for the following reasons:
(a) The constant decreases with $N$, as seen in Fig.3a. (b) The constant
depends in various ways on the shape of the contour. We checked that
by distorting the rectangle  to a slightly irregular quadrangle 
the constant
drops to zero for all $N$ as the size increases. 
In Fig3.b we show various regular polygons approximating a circle
of radius $R$:  As we increase the number of edges the constant
goes to zero for large $R$.

It is clear from the mentioned features of the Wilson loops that
there is an area law neither for very small nor very large areas.
An intermediate region in which an area law holds might still be
present.  We checked, by going to rather larger $N$, that this is
not the case for the bosonic models. There has been an interesting
suggestion \cite{poster} that such an intermediate region may exist
for the $D=4$ supersymmetric model.

\section{Conclusions and outlook}

We showed how numerical Markov chain methods can be used to verify 
{\it non-perturbative}
convergence conditions for Yang-Mills integrals with and without
supersymmetry. The same methods may be applied to establish the convergence
properties of correlation functions. Applying the technique to
invariant correlators of a single matrix, we are able to accurately
predict the asymptotic behavior of the eigenvalue density of Yang-Mills
matrix models. The results demonstrate an unusual powerlaw behavior
which, in the supersymmetric cases, persists for large $N$.
This indicates that the large $N$ limit of these ``new'' matrix models
might indeed be very different from the one of the ``old''
Wigner type models.
We also demonstrated that Monte Carlo methods are capable to rather accurately
compute various quantities relevant to these models
such as partition functions, correlation
functions, spectral distributions and Wilson loops. As opposed to
Yang-Mills quantum mechanics \cite{ch},\cite{dhn},\cite{bfss} we are confronted
to a 
system which allows some non-perturbative analysis, at least for
finite $N$.

Yang-Mills integrals are thus an
ideal laboratory for exploring new large $N$ techniques. 
Powerful analytical methods
will have to be developed if we are to verify or, maybe more importantly,
if we are to bring to good use the ideas presented in 
\cite{dhn},\cite{bfss},\cite{ikkt1}.

\section*{Acknowledgements}

This work was supported in part by the EU under Contract
FMRX-CT96-0012.


\end{document}